\documentclass[showpacs,reprint,aps, prl,longbibliography]{revtex4-1}%

\usepackage{amsmath}
\usepackage{graphicx}
\usepackage{amsfonts}
\usepackage{amssymb}
\usepackage{setspace}
\usepackage{color}
\usepackage{dcolumn}
\usepackage{bm}
\usepackage{multirow}
\usepackage{times}

\begin{document}

\title{Low-voltage broadband hybrid plasmonic-vanadium dioxide switches}
\author{Arash Joushaghani$^{1}$}
\author{Brett A. Kruger$^{1}$}
\author{Suzanne Paradis$^{2}$}
\author{David Alain$^{2}$}
\author{J. Stewart Aitchison$^{1}$}
\author{Joyce K. S. Poon$^{1}$}

\thanks{\textbf{Corresponding authors:} J. K. S. Poon and A. Joushaghani}

\affiliation{$^1$Department of Electrical and Computer Engineering and Institute for Optical Sciences, University of Toronto, 10 King's College Road, Toronto, Ontario, M5S 3G4, Canada}
\affiliation{$^2$Defence Research and Development Canada - Valcartier, 2459 Pie-XI Blvd. North, Quebec, Quebec G3J 1X5, Canada}

\date{\today }

\maketitle
\textbf{Surface plasmon polaritons can substantially reduce the sizes of optical devices, since they can concentrate light to (sub)wavelength scales \cite{Ozbay2006, MacDonald2010,Schuller2010}. However, (sub)wavelength-scale electro-optic plasmonic switches or modulators with high efficiency, low insertion loss, and high extinction ratios remain a challenge due to their small active volumes \cite{Dionne2009, Min2009, Cai2009}. Here, we use the insulator-metal phase transition of a correlated-electron material, vanadium dioxide, to overcome this limitation and demonstrate compact, broadband, and efficient plasmonic switches with integrated electrical control.  The devices are micron-scale in length and operate near a wavelength of 1550 nm. The switching bandwidths exceed 100 nm and applied voltages of only 400 mV are sufficient to attain extinction ratios in excess of 20 dB.  Our results illustrate the potential of using phase transition materials for highly efficient and ultra-compact plasmonic switches and modulators.}

The miniaturization of optical devices, whether using surface plasmon polaritons (SPPs) at metal-dielectric interfaces or strongly confined electromagnetic modes in high-index-contrast dielectric waveguides, presents an opportunity to reduce the sizes of electro-optic switches and modulators. As the volume of the active region shrinks, if the optical confinement is maintained, the amount of energy required to activate the material for modulation decreases. However, small active regions and low power consumption often come at the cost of a reduced extinction ratio. In dielectric devices, this trade-off is often overcome by recirculating light in high-$Q$ microcavities, which restrict the operation wavelengths to narrowband cavity resonances \cite{Xu2005}. Plasmonic devices can be significantly more broadband, but the losses of SPPs typically prevent high-$Q$ microcavities from being formed and limit the devices to micron or sub-micron lengths. These constraints have led to large switching voltages $ > 10$ V \cite{Dicken2008,Melikyan2011} and extinction ratios that are at best $\sim 10$ dB \cite{Volker2012,Dionne2009} for short plasmonic switches that are no more than several microns long. 

\begin{figure}[htb]
\includegraphics[width=8.6 cm]{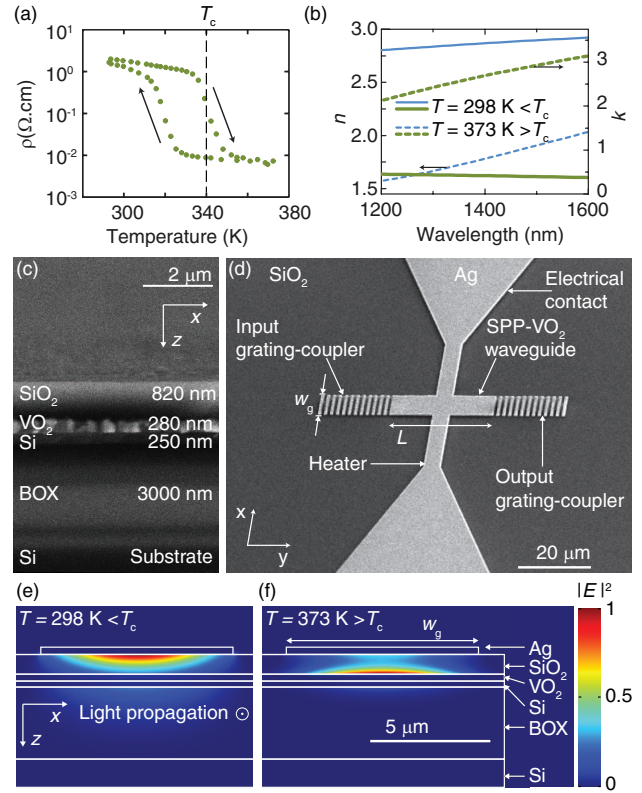}
\caption{\textbf{VO$_2$ properties and the hybrid SPP-VO$_2$ switch.} (a) Measured temperature-dependent resistivity, $\rho$, of the VO$_2$ films. The phase transition occurred at $T_c \approx 340$ K. (b) The  real, $n$, and imaginary, $k$, components of the refractive index of the deposited VO$_2$ film measured by ellipsometry. (c) SEM image of the device cross-section with the thicknesses of various layers labeled. (d) SEM image of the hybrid SPP-VO$_2$ switch with an integrated heater. The simulated electric field intensities of the electromagnetic modes of the SPP-VO$_2$ waveguide when the temperature of VO$_2$ layer is (e) $T<T_c$ and (f) $T>T_c$.}
\label{fig:fig1}
\end{figure}

Efficient and compact electro-optic plasmonic switches require materials that exhibit exceptionally large changes in their refractive indices when subjected to electrical control signals. A promising material for infrared wavelengths is the correlated-electron material, vanadium dioxide (VO$_2$). The electronic distribution and lattice of VO$_2$ can reversibly reconfigure to produce an insulator-metal phase transition with a resistivity change of up to three orders of magnitude. The phase transition can be initiated by temperature changes \cite{Berglund1969}, optical \cite{Cavalleri2001} or terahertz \cite{Liu2012} pulses, electric fields \cite{Stefanovich2000}, surface charge accumulations \cite{Nakano2012}, or mechanical strain \cite{Cao2009}.  

In this work, we demonstrate plasmonic switches with low switching voltages and record high extinction ratios using a hybrid SPP-VO$_2$ geometry and the thermally-induced VO$_2$ phase transition. In contrast to previous proposals and demonstrations \cite{Volker2012,Dionne2009,Randhawa2012,Papaioannou2012,Briggs2010,Sweatlock2012}, our devices are highly compact (between 5 to 15 $\mathrm{\mu m}$ long) and have integrated electrical control. To show the potential for integration with silicon waveguides, polycrystalline VO$_2$ was first deposited on a silicon-on-insulator (SOI) substrate using radio-frequency (RF) magnetron sputtering at a substrate temperature of 773 K (details described in Methods). The thicknesses of the VO$_2$, Si, and buried oxide (BOX) layers were 280 nm, 250 nm, and 3 $\mathrm{\mu m}$, respectively. The temperature-dependent resistivity of this film, shown in Fig. \ref{fig:fig1}(a), indicates that the critical temperature for the insulator-metal phase transition was about $T_c \approx 340$ K, typical for VO$_2$. 

Accompanied with the change in the resistivity in the phase transition is a dramatic change in the real, $n$, and imaginary, $k$, parts of the refractive index in the infrared wavelength range as shown in Fig. \ref{fig:fig1}(b) \cite{Berglund1969}. The measured refractive indices for our VO$_2$ film at $\lambda = 1550$ nm were $2.9+0.4i$ at $T  = 298$ K $< T_c$ and $2.0+3.0i$ at $T = 373$ K $> T_c$. Even though VO$_2$ exhibits a large index change near $T_c$, it is not an ideal waveguide material by itself; the imaginary part of the refractive index is too high in the insulator phase to support a low-loss optical mode, but it is not high enough in the metal phase to support a low-loss SPP. Therefore, to form a switch, we incorporated VO$_2$ in a hybrid geometry with a plasmonic waveguide \cite{Kruger2012,Sweatlock2012}.

The devices were formed in a 300 nm thick silver (Ag) film on a 820 nm thick SiO$_2$ spacer on the VO$_2$ using a lift-off step. Figures \ref{fig:fig1}(c)-(d) respectively show scanning electron microscope (SEM) images of the material stack and device. The Ag features and thickness of SiO$_2$ spacer were designed using combined optical, thermal, and electrical modelling to optimize the extinction ratio (ER), insertion loss (IL), and switching voltage, $V$, for the given thicknesses of VO$_2$ and SOI. The grating couplers (10 periods with a period of 1040 nm) coupled light into and out of the SPP-VO$_2$ waveguide. A 5 $\mathrm{\mu m}$ wide strip perpendicular to the waveguide acted as a thin film heater to heat the local volume of VO$_2$ above $T_c$ when a current passed through it. The heater widened to the contact pads.

The SPP-VO$_2$ waveguides were $w_g=8$ $\mathrm{\mu m}$ wide to facilitate the measurements and were varied in length, $L$. To achieve a high extinction ratio, the SPP-VO$_2$ waveguides should switch between a pair of hybrid modes \cite{Oulton2008,Alam2010} depending on the phase of the VO$_2$. To ensure that the switching is due to propagation through the SPP-VO$_2$ waveguides and not a change in the input/output coupling losses between the waveguide and the gratings, this pair of hybrid modes should have different propagation losses but similar effective indices. At $T < T_c$, our designed SPP-VO$_2$ waveguide supports a low-loss, transverse magnetic (TM) polarized mode shown in Fig. \ref{fig:fig1}(e), which is similar to the SPP mode of a single Ag-SiO$_2$ interface \cite{Kruger2012}. This mode has a calculated effective index of $n_\mathrm{eff} = 1.45$ and a propagation loss of 0.4 dB/$\mathrm{\mu}$m. At $T > T_c$, the waveguide supports a TM polarized metal-oxide-metal-like mode that is mainly confined between the VO$_2$ and Ag layers as shown in Fig. \ref{fig:fig1}(f). By tracing the modes while increasing thickness of the SiO$_2$ spacer layer, the hybrid mode of Fig. \ref{fig:fig1}(f) is found to be of a different order compared to the low-loss mode of Fig. \ref{fig:fig1}(e). This lossy mode has a calculated effective index of $n_\mathrm{eff} = 1.50$ and a propagation loss of 3.1 dB/$\mathrm{\mu}$m. Therefore, using the VO$_2$ phase transition, the excitation of the two different types of modes leads to a theoretical ER-per-length of 2.7 dB/$\mathrm{\mu}$m.

\begin{figure}[tb]
\includegraphics[width=8.6 cm]{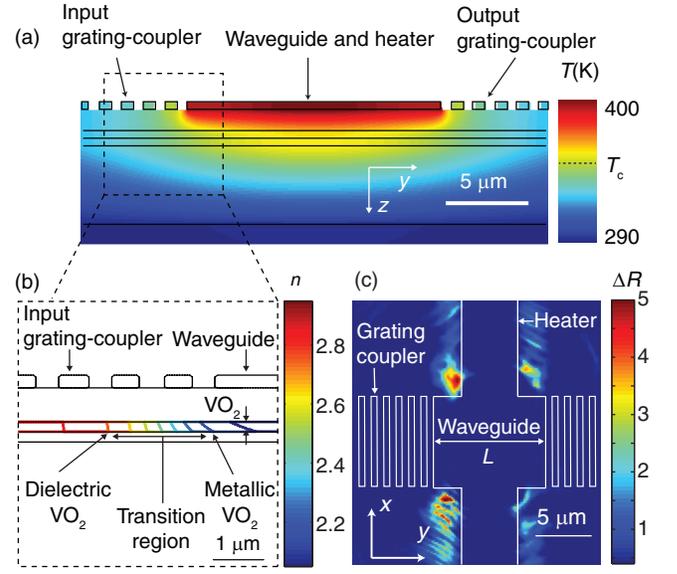}
\caption{\textbf{The temperature distribution of the device and the associated refractive index profile in the VO$_2$.} (a) The simulated temperature profile along $y$ at the mid-line of the SPP-VO$_2$ waveguide under an applied current of $I_0 = 180$ mA. (b) The refractive index contours of VO$_2$ film near the edge of the hybrid SPP-VO$_2$ waveguide computed using the temperature profile of \ref{fig:fig2}(a). The regions with insulating and metallic VO$_2$ phases are marked with arrows. The fully metallic VO$_2$ region is directly under the heater. (c) The change in the surface reflectivity, $\Delta R$, measured in  the vicinity of the heater and SPP-VO$_2$ waveguide (estimated position marked by white lines). The heater is along $x$, and the waveguide is along $y$.}
\label{fig:fig2}
\end{figure}

We designed the devices such that the heating would minimally affect input/output coupling losses, so our results can be generalized to cases with integrated input/output waveguides. Figure \ref{fig:fig2}(a) shows the temperature distribution along  the mid-line of the device computed from a three-dimensional thermal simulation with an applied current of $I = 180$ mA. The heat is localized under the SPP-VO$_2$ waveguide and, as a result, mostly changes the phase of the VO$_2$ directly under the heater. Figure \ref{fig:fig2}(b) shows the contours of the refractive index in the VO$_2$ film computed from the temperature profile of Fig. \ref{fig:fig2}(a). The VO$_2$ under the heater experiences the largest index change and is completely metallic, while the VO$_2$ under the gratings remain predominantly in the insulating phase. Since the spot size of the optical input in the experiment spanned $\sim$ 6 grating periods and was positioned near the center of the grating-coupler, for the currents tested, the input grating-coupler mainly excited the low-loss SPP-VO$_2$ mode of Fig. \ref{fig:fig1}(e). This mode propagated through the transition region of Fig. \ref{fig:fig2}(b) and would in turn be converted to the high-loss mode of Fig. \ref{fig:fig1}(f) when the VO$_2$ under the waveguide region was in its metallic phase. Electromagnetic simulations show that the theoretical net input and output coupling losses due to the grating-couplers is about 14.5 dB. Using the temperature profile of Fig. \ref{fig:fig2}(a), the net change in the coupling losses, resulting from both the change in grating coupling losses and the mismatch between the two modes of Fig. \ref{fig:fig1}(e)-(f), is estimated to be $< 3$ dB. 

To experimentally confirm that the heating was localized, we uniformly illuminated the fabricated devices from the top at a wavelength of $\lambda = 1550$ nm to image the change in the reflectivity, defined as 
\begin{equation}
\Delta R = \frac{R_{I=I_0} - R_{I=0}} {R_{I=0}},
\end{equation} 
where $R_I$ is the surface reflectivity when a current of $I$ passes through the heater. Figure \ref{fig:fig2}(c) shows $\Delta R$ at $I_0 = 180$ mA.  $\Delta R$ is localized near the edges of the heater and SPP-VO$_2$ waveguide, away from the grating-couplers. $\Delta R > 0$ is in agreement with the reflection spectra extracted from the ellipsometry data of the VO$_2$ film on SOI. 

\begin{figure}[t]
\includegraphics[width=8.6 cm]{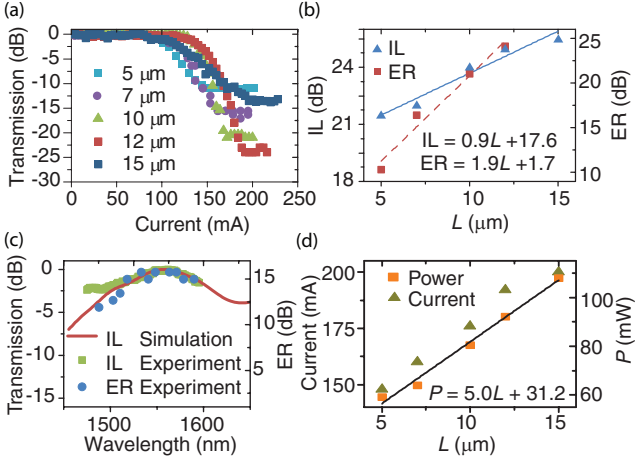}
\caption{\textbf{The static transmission and power consumption characteristics.} (a) The normalized transmission of devices with varying lengths of SPP-VO$_2$ waveguides ($L$) as a function of the input current. An ER of 24.1 dB is reached when $L = 12$ $\mathrm{\mu m}$. For $L \geq 15$  $\mathrm{\mu m}$ , the measured ER was limited by the background noise. (b) The measured IL and ER of the SPP-VO$_2$ waveguides as a function of $L$. Linear fits of IL and ER are included. (c) The spectral dependence of IL and ER of the  $L= 7$ $\mathrm{\mu}$m device. (d) The current and electrical power required for the reported ERs as a function of $L$.}
\label{fig:fig3}
\end{figure}

\begin{figure}[t]
\includegraphics[width=8.6 cm]{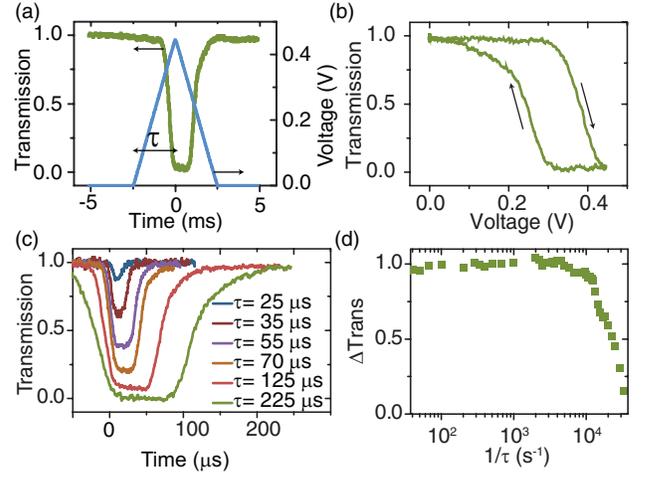}
\caption{\textbf{The dynamic switching characteristics of the $\boldsymbol{L=10}$ $\mathbf{\mu m}$ device.} (a) (right axis) The time-dependent  transmission when a (left axis) triangular voltage pulse with a ramp duration of $\tau = 2.5$ ms was applied at the heater. (b) Transmission-voltage hysteresis of (a). (c) The optical transmission when the ramp duration of the drive voltage is reduced. (d) The change in transmission, $\Delta\mathrm{Trans}$,  as a function of $1/\tau$.}
\label{fig:fig4}
\end{figure}

\begin{figure*}[t]
\includegraphics[width=17.6 cm]{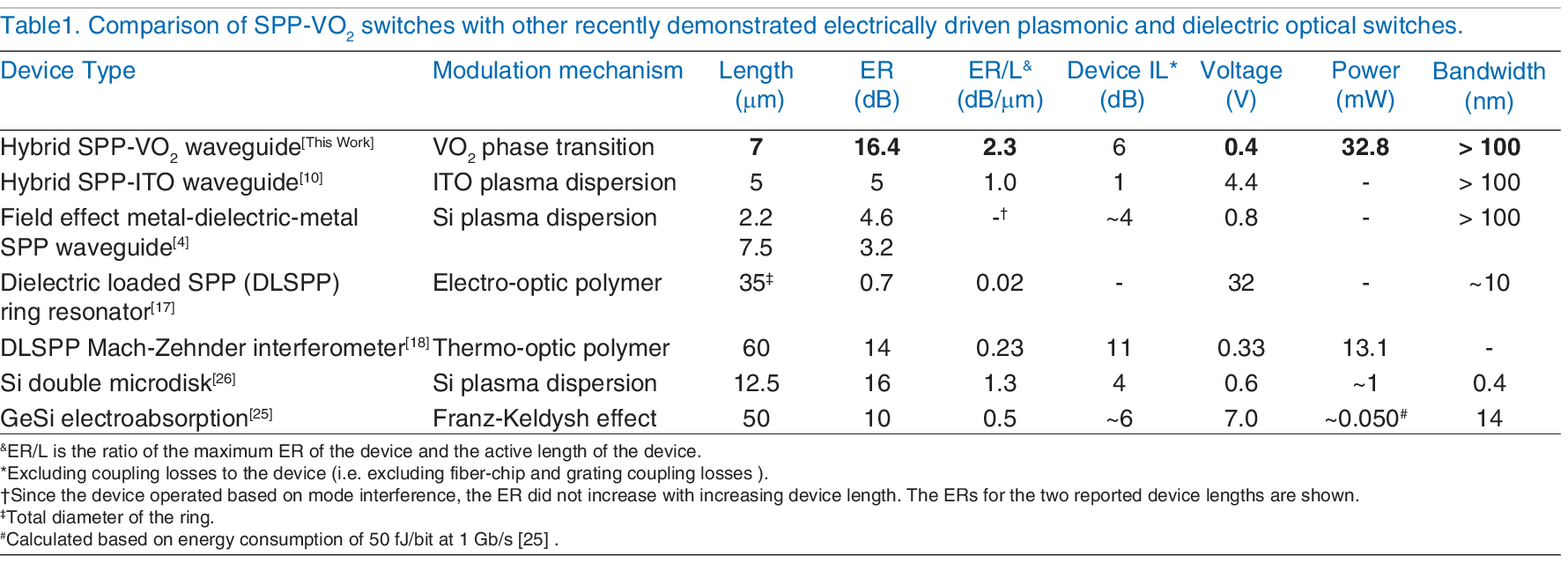}
\label{tab:tab1}
\nonumber
\end{figure*}

To investigate the device operation, we measured the static and dynamic characteristics of a set of devices with identical grating couplers and integrated heaters, but different SPP-VO$_2$ waveguide lengths between $L=5$ $\mathrm{\mu m}$ and $L=15$ $\mathrm{\mu m}$. Figure \ref{fig:fig3}(a) shows that when the applied current was increased, the normalized transmission at $\lambda = 1550$ nm dropped abruptly as $T_c$ was reached. As shown in Fig. \ref{fig:fig3}(b), the ER and IL increased with the VO$_2$-SPP waveguide length. An ER of 10.3 dB was achieved when $L=5$ $\mathrm{\mu}$m and increased to a maximum of 24.1 dB when $L=12$ $\mathrm{\mu}$m. For $ L \geq 15$ $\mathrm{\mu}$m, the optical output power became too low to measure the ER accurately. The ER of the $L=15$ $\mathrm{\mu}$m device in Fig. \ref{fig:fig3}(a) was limited by background noise. 

Linear fits of the data show that the ER-per-length was $1.9 \pm 0.2 $ dB/$\mathrm{\mu}$m and the IL-per-length was $0.9 \pm 0.1$ dB/$\mathrm{\mu}$m. This ER-per-length, to our knowledge, is the highest reported to date amongst plasmonic switches (Table 1). The intercept of the IL linear fit indicates the losses due to the grating-couplers  were  $17.6 \pm 1.1$ dB. The intercept of the ER linear fit of $1.8 \pm 2.2$ dB represents the difference between the grating-coupler losses when the VO$_2$ was in the insulator and metal phases and the mode conversion losses at the transition between the metallic and insulating VO$_2$ regions. This small offset supports that the switching was mainly due to the propagation through the SPP-VO$_2$ waveguide and not lumped coupling losses. The measured IL, ER, and grating coupling losses are in good agreement with the theoretical values from the simulations.

The broadband nature of the devices is evident by the spectra of the transmission below $T_c$ and the ER of the $L=7$ $\mathrm{\mu m}$ device in Fig. \ref{fig:fig3}(c). The measured device bandwidth was limited by the grating couplers, which had a 3 dB bandwidth of about $117 \pm 15$ nm. The calculated dispersion relation of the VO$_2$-SPP waveguide suggests the theoretical 3 dB bandwidth of the ER is about 160 nm \cite{Kruger2012}.

We estimate the power consumed by the switch by subtracting the potential difference at the contacts from the net potential difference across the probes. Figure \ref{fig:fig3}(d) shows the current and power consumption for different waveguide lengths. From the linear fit, the total power dissipated in regions of the heater away from the SPP-VO$_2$ waveguide was about $31.2 \pm 3$ mW, and the power used to switch the SPP-VO$_2$ waveguide was $5.0 \pm 0.3$ mW/$\mathrm{\mu m}$.  About 28 mW  of power was required to switch the $L = 5$ $\mathrm {\mu m}$ SPP-VO$_2$ waveguide to attain an ER of 10.3 dB. The applied current was  about 148 mA and the voltage was about 400 mV.  


Since the switching was reversible and repeatable, we investigated the modulation dynamics by driving the heater with a periodic train of triangular voltage pulses with various amplitudes and ramp times, $\tau$. The delay between the pulses was 100 ms. Figure \ref{fig:fig4}(a) shows the time-dependent optical transmission of the $L=10$ $\mathrm{\mu m}$ SPP-VO$_2$ switch under an applied voltage pulse with $\tau = 2.5$ ms. The voltage was sufficiently high for complete switching. The transmission remained at a minimum for about 1 ms after the peak of the voltage pulse because of the intrinsic hysteresis of the VO$_2$ insulator-metal transition (Fig. \ref{fig:fig1}(a)) and the finite time required for thermal dissipation. This hysteresis is evident in Fig. \ref{fig:fig4}(b), where the transmission is plotted against the applied voltage.

Next, we changed $\tau$ of the voltage pulses to measure the frequency response of the  $L = 10$ $\mathrm{\mu m}$ device. Figure \ref{fig:fig4}(c) shows the transmission at several values of $\tau$, while the voltage amplitude was kept to $V_0 = 450$ mV. As $\tau$ shortened, the ER diminished because the temperature change could no longer track the voltage pulse. Figure \ref{fig:fig4}(d) shows the change in the transmission of the switch, 
\begin{equation}
\Delta \mathrm{Trans} = \left|  \frac{\mathrm{Trans}_{V=V_0} - \mathrm{Trans}_{V=0}}{\mathrm{Trans}_{V=0}}\right|, 
\end{equation}
as a function of $1/ \tau$. The 50\% roll-off frequency was about 25 kHz, typical of thermally activated devices \cite{Murzina2012}. Increasing the applied voltage amplitude increased the roll-off frequency, but could damage the Ag strip.

Finally, we compare the hybrid SPP-VO$_2$ switch against state-of-the-art plasmonic \cite{Volker2012,Dionne2009,Randhawa2012,Papaioannou2012} and Si-based \cite{Liu2008,Watts2011} optical switches in Table 1 to illustrate its unique advantages. The $L = 7$ $\mathrm{\mu m}$ SPP-VO$_2$ switch is included in the table. The hybrid SPP-VO$_2$ switch is one of the most compact switches demonstrated to date, yet it simultaneously has the highest ER-per-length and among the lowest switching voltages. Its ER (16.4 dB) is superior to other plasmonic switches and similar to the double Si microdisk switch \cite{Watts2011}. Its switching power is of the same order of magnitude as other thermo-optic switches.  The power can be improved by shortening the heater, narrowing the waveguide, and removing the Si underneath the VO$_2$. We expect that the power consumption can be reduced to 1 - 10 mW. The Si microdisk switch \cite{Watts2011} did not have thermal tuners to bias the wavelength, which would have added about 10-20 mW of power \cite{Li2011}. The GeSi electro-absorption switch \cite{Liu2008}, though significantly larger in size, had a low power consumption because it operated by the field-induced Franz-Keldysh effect rather than a thermo-optic effect. The comparison shows that hybrid SPP-VO$_2$ switches have the unparalleled capability to exhibit large ERs at short device lengths while maintaining a broad operation bandwidth. 

In summary, we have demonstrated the first hybrid plasmonic switches that use a transition metal oxide, VO$_2$.  Our results show that  through the choice of materials and design, electrically-controlled plasmonic switches can have performance characteristics competitive and superior to their dielectric counterparts.  This work opens the avenue toward using VO$_2$ for (sub)wavelength size-scale yet efficient opto-electronic devices. The ability of the VO$_2$  phase transition to be initiated at sub-picosecond timescales by  electric fields \cite{Stefanovich2000, Nakano2012} are  promising for ultra- high-speed and low-power modulation.

\noindent\textbf{Methods:}{\footnotesize 
The devices were designed using a commercial finite element solver (COMSOL Multiphysics). The optical design was carried out using the RF module, and the electrical and thermal designs were carried out using coupled heat transport and DC current modules. The optical properties of silver and SiO$_2$ were taken from \cite{Palik1998}. The thermal properties of VO$_2$ were derived from \cite{Berglund1969,Oh2010}.

The VO$_2$ film was deposited on SOI substrates using RF magnetron sputtering at 100 W RF power and a pressure of 7 mT. The substrate was kept at a temperature of 773 K. The sputtering used a 2 inch diameter 99.7 \% vanadium target along with a gas mixture of argon (Ar) and oxygen (O$_2$) with a flow rate of 83 sccm for Ar and 4.15 sccm for O$_2$, resulting in a O$_2$ concentration of 5.77\%. The resistivity of the VO$_2$ film on SOI was measured using a four-point-probe setup (Four-Dimensions Six-Point-Probe 101C). The refractive indices were measured using ellipsometry with the sample mounted on a temperature controlled stage (Horiba Jobin Yvon UVISEL spectroscopic ellipsometer) \cite{Crunteanu2010}.

The fabrication of the devices used electron-beam lithography and lift-off. First, a 200 nm planarizing spin-on-glass (HSQ 6\%) was spin-coated and cured on top of the VO$_2$ film. A 620 nm SiO$_2$ layer was then deposited using a plasma-enhanced chemical vapor deposition system (Oxford Plasmalab). The SiO$_2$ planarized the VO$_2$ to reduce the scattering losses and set the waveguide dispersion. The features were defined using electron-beam lithography (Vistec EBPG 5000+) on a resist (ZEP-520A), followed by electron-beam evaporation of $\approx$ 0.3 nm thick chromium adhesion layer and silver.  To enhance the adhesion of the electron-beam resist, a thin layer of HMDS was spin-coated on the SiO$_2$ prior to the spin-coating of ZEP-520A.  Finally, the features were formed using a single lift-off step. 

To measure the static characteristics of each device, a 20X infinity-corrected objective lens mounted over the chip was used to focus light from a tunable laser (Agilent 81682A) on the input grating and collect the optical signal from the output grating. A half-wave plate was used to adjust the polarization of the input light. The output was separated from the input light using a beam-splitter, spatially filtered, and fiber-coupled to a power meter (Agilent 81633A). DC electrical probes contacted the pads. Current was ramped up and down using a sourcemeter (Keithley 2636A) while the optical transmission and the voltage  across the electrical probes were monitored. The chip was mounted on a temperature controlled stage maintained at 298 K. By measuring the resistance of calibration heaters of different lengths, we found the contact resistance of the probes was about $2.5\pm 1$ $\mathrm{\Omega}$.

To measure the dynamic response, periodic triangular voltage pulses from a function generator (Tektronix AFG3102) were applied across electrical contacts. The same method as above was used to couple light into and out of the SPP-VO$_2$ waveguides; however, the input light was pre-amplified using an erbium doped optical amplifier before the input grating. The output light was then coupled into a nanosecond detector (New Focus 1623). The voltage output of the photodetector was captured by an 200 MHz oscilloscope (Tektronix TDS 2022B), which also simultaneously monitored the voltage across the electrical probes.
}

\noindent\textbf{Acknowledgments:}{\footnotesize 
A.J., B.A., J.S.A., and J.K.S.P. thank the University of Toronto Emerging Communications Technology Institute for access to the cleanroom facilities.  They are also grateful for the financial support of the Natural Sciences and Engineering Research Council of Canada.
}

\end{document}